\documentclass[%
 reprint,
 amsmath,amssymb,
 aps,
]{revtex4-2}

\usepackage{graphicx}
\usepackage{dcolumn}
\usepackage{bm}
\usepackage{hyperref}

\usepackage{xcolor}
\usepackage{orcidlink}
\usepackage{comment,slashed}

\begin{document}

\preprint{APS/123-QED}

\title{Attenuation of Cosmic Ray Electron Boosted Dark Matter}

\author{Tim Herbermann\,\orcidlink{0009-0003-4771-5759}}
 \email{tim.herbermann@mpi-hd.mpg.de}
\author{Manfred Lindner}%
 \email{lindner@mpi-hd.mpg.de}
\author{Manibrata Sen\,\orcidlink{0000-0001-7948-4332}}%
 \email{manibrata.sen@mpi-hd.mpg.de}
\affiliation{%
Max-Planck-Institut für Kernphysik,
Saupfercheckweg 1, 69117 Heidelberg, Germany
}%

\date{\today}

\begin{abstract}
We consider a model of boosted dark matter (DM), where a fraction of DM is upscattered to relativistic energies by cosmic ray electrons. Such interactions responsible for boosting the DM also attenuate its flux at the Earth. Considering a simple model of constant interaction cross section, we make analytical estimates of the variation of the attenuation ceiling with the DM mass and confirm it numerically. We then extend our analysis to a $Z'$-mediated leptophilic DM model. We show that the attenuation ceiling remains nearly model-independent for DM and mediator particles heavier than the electron, challenging some previous discussions on this topic. Using the XENONnT direct detection experiment, we illustrate how constraints based on energy-dependent scattering can significantly differ from those based on an assumed constant cross section. This highlights the importance of re-evaluating these constraints in the context of specific models.
\end{abstract}

\maketitle

\textbf{\emph{Introduction -- }}
Despite the remarkable success of the Standard Model (SM) of particle physics, it only ends up explaining about $\sim 15\%$ of the total matter content of the Universe. The remaining $85\%$, accounted for by the elusive Dark Matter (DM), is an indispensable part of our Universe and is invoked to explain various unaccounted-for observations, like the rotation curves of galaxies, lensing of galaxy clusters as well as anisotropies of the cosmic microwave background (CMB). However, when it comes to the nature of DM, we are still in the dark -- decades of ongoing searches have not yet shed any light~\cite{Bertone:2016nfn, Planck:2018vyg}. 

One of the key strategies in the search for DM is through direct detection, in which a search of energy depositions of ambient DM from the Galactic halo upon interaction with nucleons in the detector is explored~\cite{Schumann:2019eaa}. Although there has not been a detection so far, recent experiments like XENON~\cite{XENON:2023cxc}, LUX-Zepelin (LZ)~\cite{LZ:2022lsv,LZ:2022ysc} and PandaX~\cite{PandaX:2022ood} continue to push the detection limits. The detection prospects for these nuclear recoil searches are severely limited by the kinematics of energy deposition for sub-GeV low-mass DM~\cite{Essig:2022dfa}. To gain sensitivity to lower mass regimes, electron-DM scattering has been adopted as an additional detection channel and is being searched to probe low-mass DM~\cite{XENON:2022ltv,DAMIC-M:2023gxo,SENSEI:2023zdf,SuperCDMS:2020ymb}.

A lucrative idea to overcome the low DM mass suppression in current experiments is to search for signals from boosted dark matter (BDM). The idea is simple -- if by some mechanism, a subdominant fraction of DM is boosted to (semi-)relativistic energies, the kinematic suppression of the detection signal can be avoided. The proposal to consider upscattering by charged cosmic rays (CR) has been extensively studied in the literature ~\cite{Bringmann:2018cvk, Ema:2018bih, Cappiello:2018hsu, Dent:2019krz, Wang:2019jtk, Alvey:2019zaa,Guo:2020oum,Guo_2020a,Dent:2020syp,Xia:2021vbz,Bell:2021xff, Bardhan:2022bdg,Maity:2022exk,Guha:2024mjr,Bell:2023sdq,Cappiello:2024acu}. Additionally, new mechanisms for boosting the DM candidate have been explored, which involve upscattering by neutrinos, inelastic DM, decays of heavier components of a multi-component dark sector, and so on~\cite{Agashe:2014yua,Berger:2014sqa, An:2017ojc,Yin:2018yjn,DeRocco:2019jti, Emken:2018run,Chen:2020gcl,An:2021qdl,Das:2021lcr,Jho:2021rmn,Granelli:2022ysi,Emken:2021lgc,DeRomeri:2023ytt, Xia:2024ryt,Das:2024ghw,Lu:2023aar,Chen:2021ifo}. 

Searches for upscattered DM have the distinct advantage of not requiring additional assumptions about new interactions between the SM and DM. The interactions considered for direct detection naturally result in the formation of a population of BDM through scattering with CR electrons and nucleons. Consequently, a subdominant composition of BDM is inevitably present. However, detecting this boosted component of DM is challenging because the additional powers of the interaction coupling necessary for the boosting process also end up suppressing it. Furthermore, the BDM spectrum is also attenuated due to interaction with nucleons and electrons in the Earth.

To derive constraints on DM-nucleon/electron scattering cross sections arising out of CR-BDM, some of the earlier works focused on a constant energy-independent cross section model~\cite{Bringmann:2018cvk,Ema:2018bih,Cappiello:2018hsu,Maity:2022exk}. This assumption of an effective cross section need not hold for the entire energy range of scattering, thereby necessitating the requirement of a \emph{model-dependent} treatment ~\cite{Bardhan:2022bdg,Bell:2023sdq}. Furthermore, the attenuation treatment in these studies is usually very complicated and requires a numerical solution for a detailed analysis. In fact, for a realistic estimate with correct treatment of multiple scatterings, deflections and backscattering, one needs to use Monte-Carlo simulations~\cite{Emken:2018run,Cappiello:2023hza,Xia:2021vbz,PandaX:2023tfq,PhysRevD.100.103011} (however, see~\cite{Cappiello:2023hza} for an analytical comparison). In~\cite{Guha:2024mjr}, a semi-analytical prescription was presented without resorting to a full Monte-Carlo treatment of the attenuation effect. The constraints derived predicted the presence of an attenuation ceiling dependent on the mass of the mediator, even for large DM masses. Moreover, for energy-independent scattering, the attenuation ceiling, derived from~\cite{Bardhan:2022bdg} is shown to be a constant and independent of the DM and the mediator masses.

In this paper, we reconsider the scenario of energy-dependent upscattering of DM by CR electrons by revisiting the treatment of attenuation of BDM in the mean energy loss approximation without introducting additional approximations beyond this framework.
Under the assumption of a constant interaction cross section, we first perform analytical estimates of the effect of attenuation on BDM, which we confirm numerically. We demonstrate analytically the presence of an almost \emph{model-independent} attenuation ceiling for large DM masses, in contrast to the findings presented in the previous studies mentioned above. We point out the inaccuracies of the some of the approximations considered in previous studies. We further generalise our results to include energy-dependent scattering for a $Z'$ mediated leptophilic DM model and confirm that the attenuation ceiling depends only mildly on the mediator mass. We perform a detailed study of how XENONnT can constrain such models of BDM and demonstrate that constraints for an energy-dependent scattering can differ considerably from those derived under the assumptions of a constant cross section.

\textbf{\emph{Cosmic ray electron upscattered dark matter -- }}
Energetic CR electrons scattering off the ambient DM can boost the latter to kinetic energies significantly exceeding that from virial motion. The kinetic energy of the DM of mass $m_\chi$ is given by 
\begin{equation}
\begin{aligned}
    T_\chi &= T_\chi^{\text{max}} \left(\frac{1-\cos\theta}{2}\right)\,,\\
    T_\chi^{\text{max}} &= \frac{T_e^2 + 2m_e T_e}{T_e + (m_\chi+m_e)^2/2m_\chi}\,.
    \label{eq:kinematics}
\end{aligned}
\end{equation}
Here $\theta$ is the scattering angle in the centre of momentum frame, while $T_e$ and $m_e$ denote the CR electron kinetic energy in the laboratory frame and its mass respectively. The maximum energy that DM can gain in the laboratory frame is $T_\chi^{\text{max}}$.

The upscattered DM flux is given by the local upscattering rate from CR electrons, weighted by the ambient DM density and reads
\begin{equation}
    \frac{d\Phi_\chi}{dT_\chi} =  
    D_{\text{eff}}\,\rho_{\chi,\text{loc}} \int_{T_e^{\text{min}}}^{\infty} dT_e \frac{1}{m_\chi} \frac{d\Phi_e}{dT_e} \frac{d\sigma_{e\chi}}{dT_\chi} \,.
    \label{eq:upscatterd_flux}
\end{equation}
Here $\rho_{\chi,\text{loc}}$ is the local energy density of DM, which we fix to $0.4\,\mathrm{GeV\,cm^{-3}}$ and $D_{\text{eff}}$ is the effective distance to which we include CR electron boosted DM. We use $D_{\text{eff}} = 1\,\mathrm{kpc}$, only including contributions from our galactic neighborhood. This is conservative since the CR electron flux is only known locally and is subject to considerable astrophysical uncertainties. The minimum electron energy $T_e^{\text{min}}$ to accelerate DM to a kinetic energy of $T_\chi$ follows directly from kinematic considerations and can be calculated from inverting the expression for $T_\chi^{\text{max}}$ in Eq.\,\ref{eq:kinematics}.

\begin{figure}[!t]
    \centering
    \includegraphics[width=\linewidth]{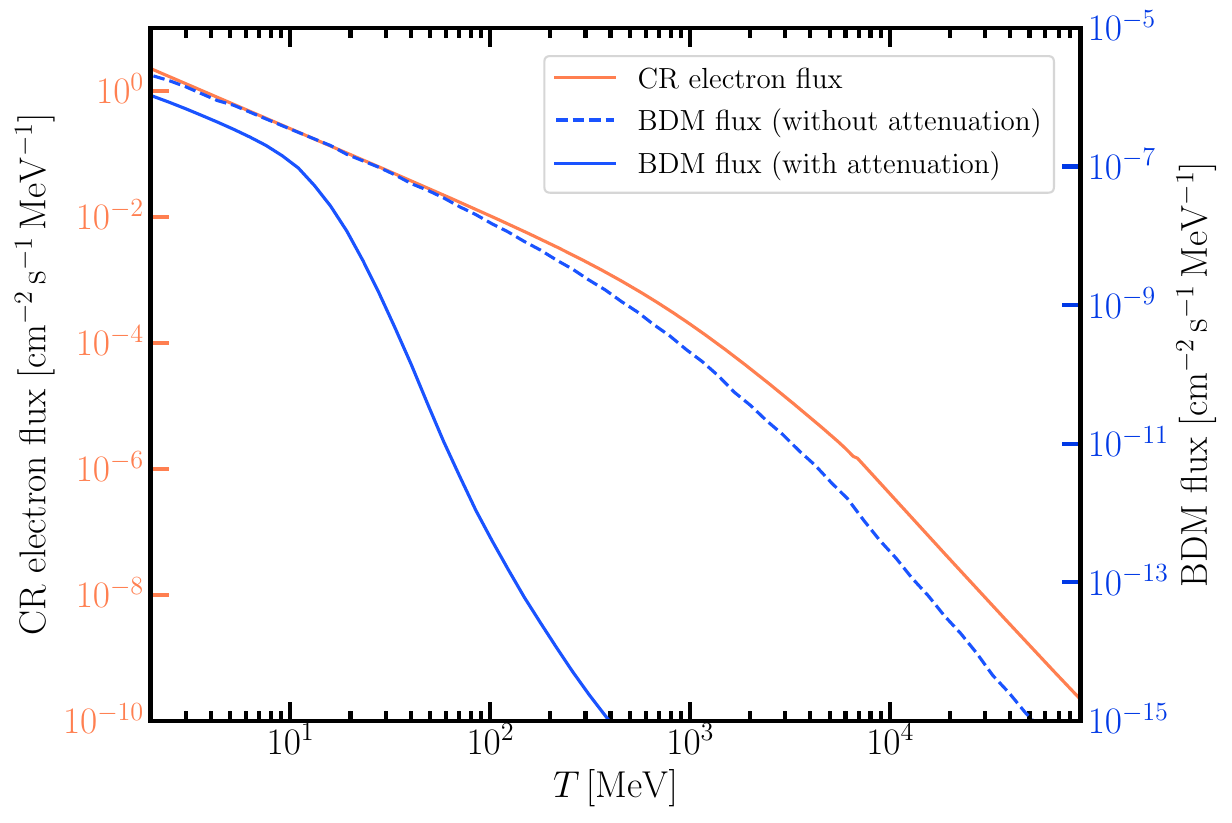}
    \caption{The CR electron flux (in red) used to boost DM and the resulting spectrum of BDM flux (in blue). For the BDM computation, we assume a constant cross section for a benchmark point $m_\chi = 10\mathrm{\,MeV}$ and $\bar\sigma_{e\chi} = 10^{-29}\mathrm{\,cm^2}$. We show the BDM flux with (blue,solid) and without (blue, dashed) attenuation of DM flux in the Earth to highlight the importance of attenuation. Note the two different value ranges for the respective axes.}
    \label{fig:CR-DM_Spectra}
\end{figure}

To perform our computations, we use an analytic parameterization of the local CR electron spectrum from~\cite{Boschini:2018zdv}. The spectra is presented from $2\,\mathrm{MeV}$ to $90\,\mathrm{GeV}$ 
%
%
%
%
and we extrapolate to lower energies, if necessary. 
The CR electron spectrum, shown in Fig.\,\ref{fig:CR-DM_Spectra}, demonstrates the well-known power-law feature.

Eq.\,\ref{eq:upscatterd_flux} can be solved numerically under the assumption of an effective constant cross section, such that $d\sigma_{e\chi}/dT_{\chi} = \bar{\sigma}_{e\chi}/T_{\chi}^{\rm max}$.
As a demonstration, an example BDM spectrum is shown in blue dashed lines in Fig.\,\ref{fig:CR-DM_Spectra} for a benchmark point $m_\chi = 10\mathrm{\,MeV}$ and  $\bar\sigma_{e\chi} = 10^{-29}\mathrm{\,cm^2}$. This demonstrates that the BDM spectra, in absence of attenuation, can have a shape similar to that of the CR electrons, albeit with a very different normalization. We expect this to change when attenuation is included, which we will describe next.

\textbf{\emph{Attenuation in the Earth -- }}
To model the energy loss of DM traversing a medium we adopt the widely used mean energy loss equation \cite{Bringmann:2018cvk,Ema:2018bih,Das:2024ghw,Dent:2019krz}
\begin{equation}
    \frac{dT_\chi}{dx}(x) = - n_e \int_{0}^{T_e^{\text{max}}} dT_e \, T_e \frac{d\sigma_{e\chi}}{dT_e}\,,
    \label{eq:energyloss}
\end{equation}
where $T_e^{\text{max}}$ is the maximum energy of recoil, evaluated in a manner similar to Eq.\,\ref{eq:kinematics}
We fix the electron density in the Earth to $n_e = 8\times 10^{23} \,\mathrm{cm^{-3}}$ \cite{DeRomeri:2023ytt}.

For $T_\chi \ll m_e$ and assuming $d\sigma_{e\chi}/dT_{e} = \bar{\sigma}_{e\chi}/T_{e}^{\rm max}$,  Eq.\,\ref{eq:energyloss} can be solved analytically to estimate the effect of attenuation.

However, if the target particles are electrons, the validity of the above approximation is not guaranteed and the hierarchy of energy scales in the problem is more subtle. To shed light on this issue, we consider the full energy dependence of cross sections for specific models and resort to a numerical solution. We adopt a framework similar to the study on DSNB-boosted DM, by two of the authors, which solved Eq.\,\ref{eq:energyloss} numerically without additional approximations beyond the mean energy loss framework ~\cite{Das:2024ghw}.
There we solved \ref{eq:energyloss} for different zenith angles of incoming DM, since the overburden traversed depends on the zenith angle (for more details on the geometry, see also Appendix A in \cite{DeRomeri:2023ytt}). The flux at the detector site is given by the integrated flux from all directions.

The solid blue line in Fig.\,\ref{fig:CR-DM_Spectra} shows the BDM spectrum after attenuation for a detector at depth $d=1.4\mathrm{\,km}$. We assume a constant cross section and resort to solving energy loss numerically as described above. It is evident from Fig.\,\ref{fig:CR-DM_Spectra} how the approximation $T_\chi \ll m_e$ is not valid. Thus, a full numerical solution is required.
The inclusion of attenuation significantly alters the BDM flux at the detector site. We observe an overall reduction of flux for all energies, but it is particularly pronounced for large kinetic energies where a complete loss of BDM flux is present.

While a simple analytical solution breaks down in some cases, we can still consider it to gain some intuition on the general behaviour of the numerical solution in different regimes. To this end, we solve Eq.\,\ref{eq:energyloss} for an energy independent cross section $\bar\sigma_{e\chi}$ and find
\begin{equation}
        \frac{dT_\chi}{dx} = - \frac{1}{2} n_e \bar\sigma_{e\chi} \frac{T_\chi^2+2m_\chi T_\chi}{T_\chi + \frac{(m_e+m_\chi)^2}{2m_e}} \,.
\end{equation}
To study the behaviour of the mean energy loss, we consider the heavy DM regime ($m_\chi \gg m_e$), the intermediate regime ($m_\chi \sim m_e$), and the light DM regime ($m_\chi \ll m_e$).

\paragraph{Heavy DM regime.}
We consider the heavy DM limit, i.e. $m_\chi \gg m_e, T_\chi$, where the inequality with respect to $T_\chi$ is understood to be valid for the bulk of boosted DM flux that gives a direct detection signal and not necessarily for the entire BDM spectrum. In this case, mimicking the approximation previously made for energy loss due to scattering on nucleons (e.g.\cite{Bringmann:2018cvk}), we find
\begin{equation}
    \frac{dT_\chi}{dx} = -\frac{1}{2m_\chi \lambda_{\text{eff}}} (T_\chi^2+2m_\chi T_\chi)\,,
\end{equation}
with the effective length scale of energy loss $\lambda_{\text{eff}}^{-1} = 4\,n_e \bar\sigma_{e\chi} m_e m_\chi/(m_e+m_\chi)^2$. A suitable criterion for attenuation is $d/\lambda_{\text{eff}} \approx 1$ for a detector at depth $d$. This leads to a solution where effective cross section for attenuation scales as $\bar\sigma_{e\chi} \propto m_\chi$.

\begin{figure}[!t]
    \centering
    \includegraphics[width=\linewidth]{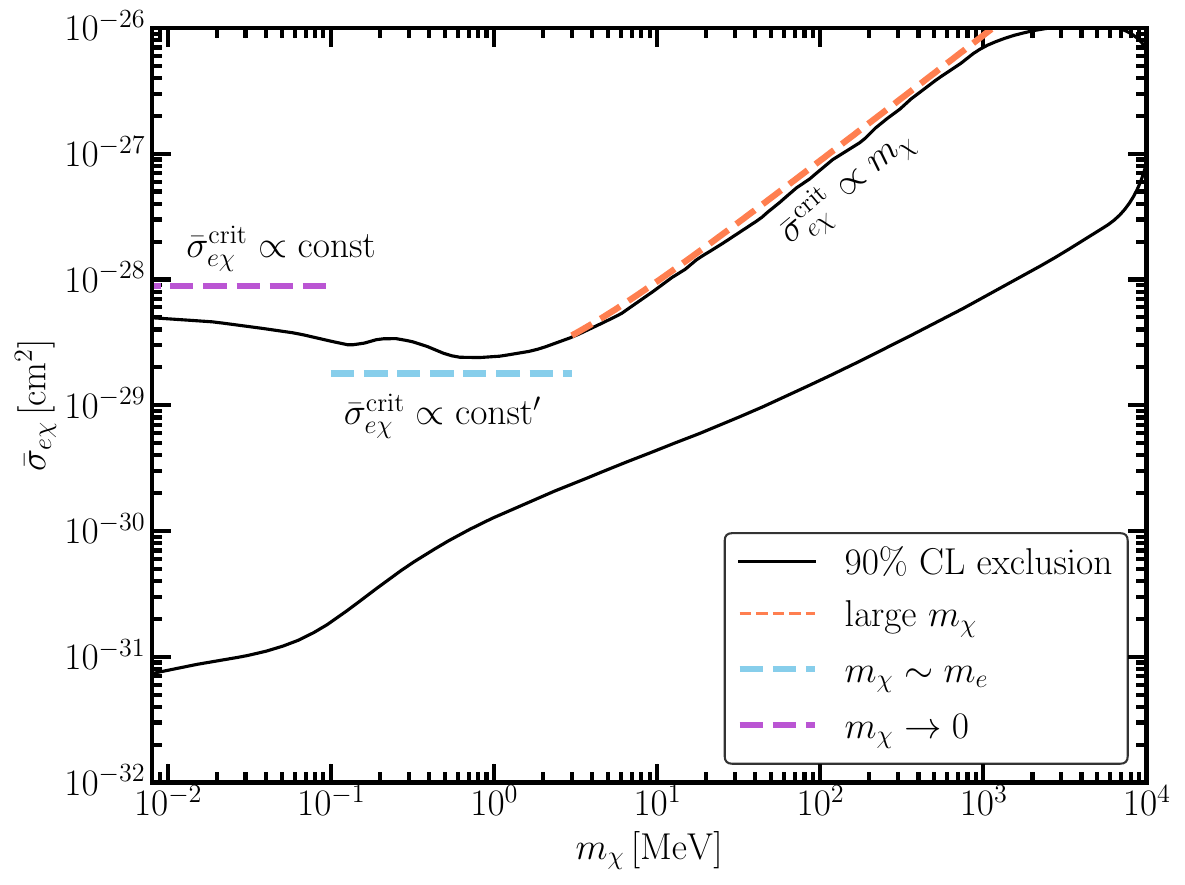}
    \caption{Numerically determined exclusion contour and attenuation ceiling for constant cross section scenario compared to the approximate analytical solutions in the small, intermediate and large $m_\chi$ limit, in the $m_\chi-\bar\sigma_{e\chi}$ plane. For heavy $m_\chi$, we note a good agreement with the numerical result, whereas for lighter results we cannot predict the precise value but are limited to qualitative statements on the scaling with $m_\chi$. }
    \label{fig:ceiling}
\end{figure}

\paragraph{Intermediate regime.}
The location of the attenuation ceiling can be estimated in the regime $m_\chi \sim m_e$. Setting $m_\chi \sim m_e \sim m$ simplifies the energy loss equation to
\begin{equation}
     \frac{dT_\chi}{dx} = - \frac{1}{2} n_e \bar\sigma_{e\chi} T_\chi \left(\frac{T_\chi + 2m}{T_\chi+2m}\right) \approx - \frac{1}{2} n_e \bar\sigma_{e\chi} T_\chi\,.
\end{equation}
This gives rise to an exponential suppression with a characteristic scale $\lambda_{\text{eff}}^{-1} =  n_e \bar\sigma_{e\chi}/2$. Using the same argument as before for a detector at depth $d$, we find a constant attenuation ceiling at $\bar\sigma_{e\chi} =2/(n_e d)$.
\paragraph{Light DM regime.}
Going to even lower DM masses, $m_\chi \ll m_e$, we infer from $T_\chi^{\text{max}}$ in Eq.\,\ref{eq:kinematics} that $T_\chi$ scales as $T_\chi \sim 2 m_\chi T_{e,\mathrm{CR}}^2/(2m_\chi T_{e,\mathrm{CR}}+m_e^2)$, where $T_{e,\mathrm{CR}}^2\gg m_e$ is the CR electron kinetic energy. Since CR electrons are relativistic, $T_\chi \gg m_\chi$ holds for most of the energy range. Thus the energy loss equation approximates to
\begin{equation}
    \frac{dT_\chi}{dx} =   - \frac{n_e \bar\sigma_{e\chi}}{2} \frac{T_\chi^2}{T_\chi+\frac{m_e}{2}}\,. 
\end{equation}
An analytic solution can be written in terms of the Lambert-$W$ function as a function of $\lambda_{\text{eff}}^{-1} = n_e \bar\sigma_{e\chi}$. We note that the precise scaling of energy loss is non-trivial and depends on the constant of integration, i.e., on the initial kinetic energy of the BDM particle. So while we cannot reliably predict the precise value, we may still infer that the location should be largely independent of $m_\chi$. Hence, we expect again a constant attenuation ceiling, albeit at different numerical values of $\bar\sigma_{e\chi}$ due to the potentially much weaker suppression. These analytical estimates allow us to understand the dynamics of attenuation of BDM due to the scattering by electrons in the atmosphere and the earth. As seen from Fig.\,\ref{fig:ceiling}, the analytical estimates give an excellent prediction of the behavior of the attenuation ceiling obtained by solving Eq.\,\ref{eq:energyloss} numerically. Some caution is warranted for regime of $m_\chi \lesssim m_e$. While the analytic solution does capture the numerical result qualitatively, it is the mean average loss approach itself that faces some limitations here.

Although these estimates are performed for a constant cross section, we expect the conclusion to qualitatively hold for fully energy-dependent cross sections as well.
To understand this, we can decompose the integral in Eq\,\ref{eq:energyloss} as $\frac{1}{2} T_e^{\text{max}} \bar\sigma_{e\chi} \times f(T_\chi)$, where $\frac{d\sigma_{e\chi}}{dT_e} = \frac{\bar\sigma_{e\chi}}{T_e^{\text{max}}}$ and $f(T_\chi)$ is an effective form factor that encodes deviations from the constant cross section assumption.
Note that both the effective cross section as well as the effective form factor have an explicit model dependence. Hence, the attenuation is well described by a constant cross section, $\bar\sigma_{e\chi}$, modulated by a model-dependence coming from $f(T_\chi)$. We find that as long as $f(T_\chi)\sim 1$, the general properties, including the position of the attenuation ceiling, are reproduced and independent of the details of the underlying interaction.

\textbf{\emph{Event rates and statistical analysis -- }}
The non-observation of BDM event rates in direct detection experiments can be used to constrain DM-electron scatterings. In this work, we restrict our analysis to electron recoil searches by the XENON collaboration \cite{XENON:2022ltv}. We expect from our previous studies of supernova neutrino background-BDM that the constraints arising from other direct detection experiments like LZ \cite{LZ:2022lsv} and PandaX \cite{PandaX:2022ood} will be qualitatively similar, although some differences may arise due to the different overburden.

The predicted event rate is calculated as 
\begin{equation}
    \frac{dR}{dT_e} = N_e \int dT_\chi \frac{d\Phi_\chi}{dT_\chi^x} \frac{d\sigma_{e\chi}}{dT_e}\,,
    \label{eq:event_rate}
\end{equation}
where $T_\chi^x$ is the solution of Eq.\,\ref{eq:energyloss}, giving the DM kinetic energy loss as a function of depth, and
$N_e =Z_\text{eff}(T_e) M_\text{det}/m_\text{Xe} $ denotes the number of electron targets in the detector. Typically, the effective charge number $Z_\text{eff}(T_e)$ is energy-dependent because of the need to overcome the electron binding energy to achieve ionization. Here we assume a constant $Z_\text{eff}(T_e) \approx 40$ \cite{Fornal:2020npv}. To calculate the flux of BDM ${d\Phi_\chi}/{dT_\chi^x}$, a detector depth of $h_d=1.4\mathrm{\,km}$ is assumed. We take the convolution of the rate with a Gaussian resolution function centred around the reconstructed energy $E_R$ and use a resolution of $\sigma_\text{XE} = 0.31 \sqrt{E/\text{keV}} + 0.0037 E/\text{keV}$\,\cite{Aprile_2020}. The convoluted signal is multiplied by the respective efficiency function.

To derive constraints on the DM mass and the couplings of the model, we employ the following $\chi^2$ statistic, 
\begin{equation}
    \chi^2 = \sum_{E_i} \frac{\left(R_i^\text{pred} - R_i^\text{exp}\right)^2}{\sigma_i^2}\,.
\end{equation}
Here $R_i^\text{pred}$ is the predicted event rate from the best-fit background model and the additional CR boosted DM contribution. Measured event rates are indicated as $R_i^\text{exp}$. We use published experimental uncertainties $(\sigma_{\text{Di}})$ and combine those with a Poissonian counting error on the total predicted event rate, i.e.
$\sigma_i^2=  R_i^\text{pred} + \sigma_{\text{Di}}^2\,$.
To exclude regions of parameter space, we use a $\chi^2$ difference to the best-fit background model, i.e. $\Delta \chi^2 = \chi^2 - \chi^2_\text{bkg}$ and use $\Delta \chi^2 > 4.61$ to reject at $90\%$ confidence level (CL). Any constraints put from this statistical model are conservative, as in they do not attempt a joint fit of background and signal which would require a more involved statistical model.

We first demonstrate our constraints for the case of the constant effective cross section, i.e.,  $d\sigma_{e\chi}/dT_{\chi,e} = \bar{\sigma}_{e\chi}/T_{\chi,e}^{\rm max}$, in a XENON-like setup in Fig.\,\ref{fig:ceiling}. We compare the numerically determined exclusion contour with the analytic estimates for the attenuation ceiling. The limit $m_\chi \gg m_e$ is captured perfectly by the analytical approximation - the numerical solution exhibits approximately the same exponential energy loss that determines the attenuation limit.

In the regime $m_\chi \sim m_e$, the approximations ensure a qualitative agreement, predicting the turnover of the attenuation ceiling. This also demonstrates that the position of the attenuation ceiling is largely independent of $m_\chi$ in this regime. The limit $m_\chi \ll m_e$, however, is less well determined. The reason is two-fold:  There is an explicit dependence on the initial energy of the particle and energy loss from attenuation is no longer exponential but rather follows a complicated functional dependence that is much weaker. However, we can still predict qualitatively a ceiling that is, within the scope of the approximations made, largely independent of $m_\chi$, a behaviour that we also find in the full numerical solution which exhibits a very weak dependence on $m_\chi$ in this regime. While the last two regimes provide valuable insight into the behaviour of attenuation ceilings that arise from the mean energy loss approximation, it should be noted that the framework itself is limited in applicability here, since DM is light compared to electrons. Thus, effects such as angular deflection or backscattering become more important but are not captured in this approach. Since these effects tend to make energy loss more efficient, we suspect that the actual attenuation ceiling is likely stronger in this regime.
 
We now turn to demonstrate the model-dependence of the results, through a model of DM-electron interaction through a massive vector gauge boson. The SM can be extended with a singlet Dirac DM $\chi$ coupled to a massive vector boson $Z_{\mu}'$. This can be embedded in the SM with an additional gauge U$(1)$ symmetry, like $L_e-L_\mu$, $L_e-L_\tau$ or a flavour universal one~\cite{Bauer:2018onh,Barman:2024lxy}. We remain agnostic of the details of the complete model. The relevant interactions for boosting and detecting the DM are
\begin{equation}
\mathcal{L} \supset g_e\,\bar{e} \gamma^{\mu} e Z_{\mu}' + g_\chi\, \bar{\chi} \gamma^\mu \chi Z_{\mu}'\,.
\end{equation}
Note that in this case, the effective cross section becomes model-dependent and is defined through,
\begin{equation}
    \bar{\sigma}_{e\chi} = \frac{g^4}{\pi} \frac{\mu_{e\chi}^2}{(q_\text{ref}^2+m_{Z'}^2)^2}\,,
    \label{eq:sigmadef}
\end{equation}
where $\mu_{e\chi}$ is the reduced mass of the electron-DM-system and $g=\sqrt{g_e g_\chi}$ is an effective coupling. The conventional definition of $q_\text{ref} = \alpha m_e$ involves the fine-structure constant. Using this parameterization allows for easy comparison to previous studies.

\begin{figure}[!t]
    \centering
    \includegraphics[width=\linewidth]{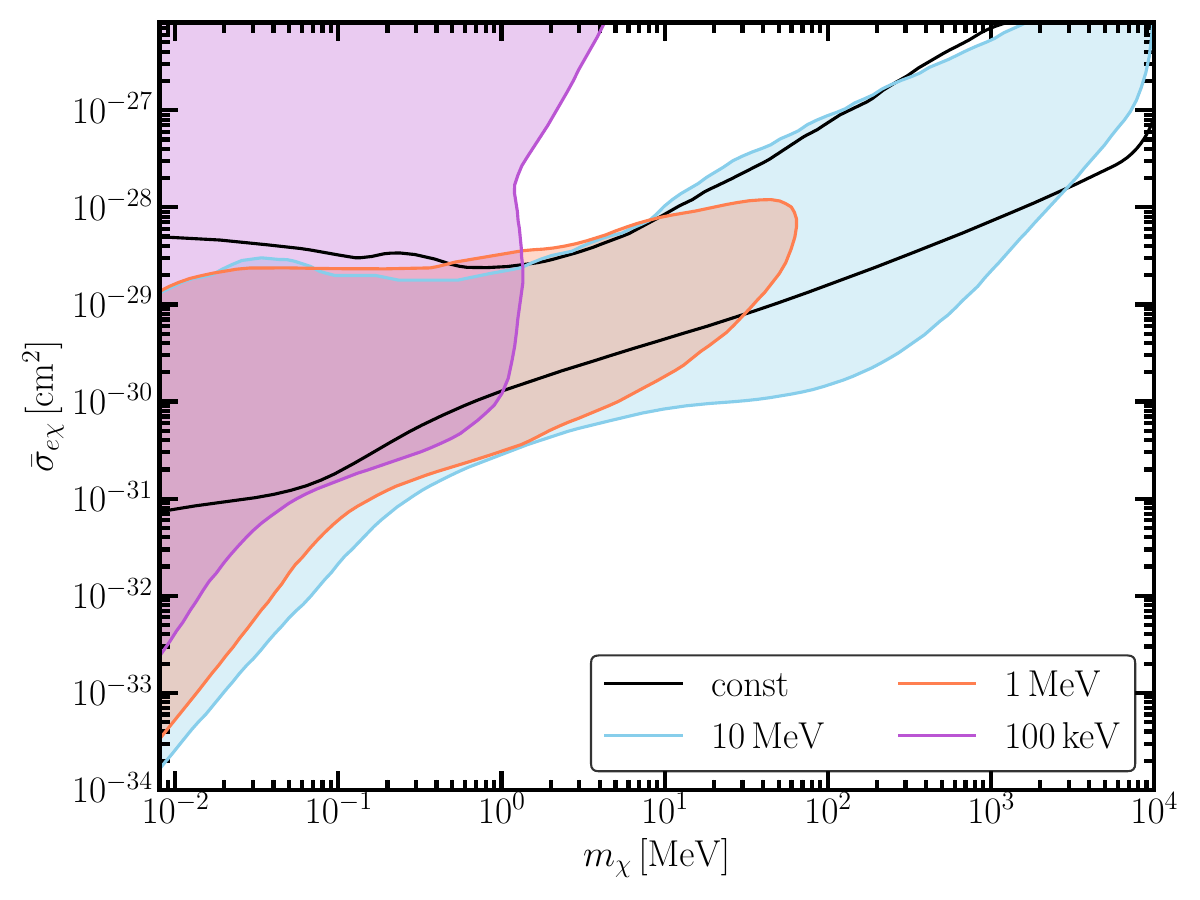}
    \caption{Constraints cast on the parameter space for our $Z'$-model. The coloured shaded regions show limits at $90\%$ CL for different masses of the mediator and present results in terms of DM mass and the effective cross section $m_\chi-\bar\sigma_{e\chi}$ plane to allow comparison to previous results.  The constraint from a constant cross section calculation is shown in black. }
    \label{fig:constraints}
\end{figure}

Fig.\,\ref{fig:constraints} shows the constraints on the $m_\chi-\bar\sigma_{e\chi}$ plane for different $Z'$ masses. To compare our results, we show the constraints on the constant cross section model in black.
As discussed before, the constraints are strongly dependent on the mediator mass and can change drastically for a low-mass mediator. Even when the mediator is heavy, the constraint can be quite different from the constant cross section case, especially for low DM masses. In light of our discussion on the approximate solutions, we can attempt to understand the general behaviour of the exclusion contours and especially the attenuation ceiling.

In the limit of large $m_\chi$ and $m_{Z'}$, the attenuation ceiling is universal and model-independent. In this limit, the integral part of the energy loss equation (Eq.\,\ref{eq:energyloss}) is well approximated by the same integral assuming the effective cross section (Eq.\,\ref{eq:sigmadef}) the model is mapped on. Since the energy loss is an integrated effect, we are less sensitive to explicit energy dependence. Towards smaller $m_\chi$ the attenuation part of the contour becomes increasingly more model-dependent and moderate differences between mediator masses are observable. This is largely driven by the effective $\bar\sigma_{e\chi}$ becoming a poorer approximation to the actual cross section.  For very small DM masses, the attenuation ceiling falls off strongly with decreasing mass. This effect is an artefact of the definition of $\bar\sigma_{e\chi}$. Indeed, as $m_\chi \rightarrow 0$ we find $\bar\sigma_{e\chi} \propto m_\chi^2$. While we need to keep in mind that our treatment of attenuation has limited applicability in this low DM mass regime, our result hightlights once more that the details of energy-dependence play a large role in making predictions and casting bounds.

For light mediators, $m_{Z'} < m_e$, the obtained contours are significantly different. This is primarily because the light mediator regime is qualitatively very different from the heavy mediator and constant cross section cases in terms of the differential cross section, as mediator masses start becoming relevant. Moreover, the commonly adapted parameterization is also a very poor proxy for the actual cross section in light mediator models of BDM. This highlights once more the necessity of model-dependent studies in the direct detection of BDM, in tandem with what was presented in~\cite{Bell:2023sdq}.

\textbf{\emph{Conclusion --}}
Boosted Dark Matter (BDM) offers a way to derive complementary constraints on Dark Matter (DM) models. When a model assumes certain interactions that permit direct detection searches, an irreducible background of BDM is almost inevitable, provided there is a suitable upscattering mechanism. In this study, we examined the upscattering of light leptophilic DM by cosmic ray electrons. By focusing on a minimal DM model where interactions are mediated by a massive vector mediator, we obtained updated constraints on DM-electron interactions using data from XENONnT, covering the mass range from $10\mathrm{\,keV}$ to $10\mathrm{\,GeV}$. We emphasise the strong model dependence of experimental signatures and constraints in BDM studies, particularly exploring the attenuation of the BDM flux in the medium before it reaches underground detectors.

Regarding the attenuation of BDM in the Earth, we considered mean energy loss of BDM from scattering on electrons in the Earth. A fluctuation of energy loss due fluctuations in the number of scattering events is neglected. Moreover, it should be noted that limitations of the mean energy loss or ballistic approach have been studied in the literature, i.e. \cite{Xia:2021vbz,PhysRevD.100.103011} for the case of relativistic BDM or \cite{Emken:2018run,Emken:2019tni} for non-relativistic halo DM.

We would like to emphasise here that our improved solution for the ballistic approximation cannot replace a full Monte-Carlo treatment of attenuation. It rather shows that this widely used straightforward approach in modelling attenuation needs to be treated correctly in order to get results that are consistent within the mean energy loss framework made. We also note that the agreement of mean energy loss and full Monte-Carlo implementations improves for DM that is substantially heavier than the scattering target in attenuation, since typical angular deflections are small and fluctuations in number of scattering events are suppressed. Thus we expect our results to be particularly robust in the regime $m_\chi \ll m_e$.

Firstly, under the assumption of a constant interaction cross section, we provide analytic expressions for the effect of attenuation on the BDM flux. We further confirmed through a detailed numerical solution the excellent agreement of our analytic approximations with the full solution. 

We used our understanding of the constant cross section analysis to generalise to the case with energy-dependent interaction cross sections. 
Under the assumption of vector-mediated interaction between DM and electron, we demonstrated how the constraints imposed from a direct detection experiment like XENONnT can be very different for energy-dependent scattering as opposed to the constraints derived from an effective constant cross section interaction.

Compared to similar previous studies~\cite{Bardhan:2022bdg,Guha:2024mjr}, we note significant qualitative and quantitative deviations, especially when including the effect of attenuation.
In particular, we showed the presence of a model-independent attenuation ceiling in the presence of large DM and mediator masses. These previous studies adopted different approximation schemes, for example, considerations of critical energy loss~\cite{Bardhan:2022bdg} or including multiple scatterings~\cite{Guha:2024mjr}. 
In our study, even though we omit deflections and the possibility of back-scattering into the atmosphere, the effect of attenuation is stronger.

Since our assumptions are conservative and minimal, we expect our results for energy loss of BDM to be robust in the limit of large DM masses and compared to other implementations of mean energy loss. Our results suggest that combined study of energy dependent interactions and a full Monte-Carlo treatment of attenuation is required to produce realiable limits for leptophilic BDM, which we leave for future work.

In light of the absence of an unequivocal direct detection of dark matter, our work highlights the importance of addressing new approaches to its direct detection. Boosted dark matter sourced from upscattering by energetic particles, promises to be a powerful tool to probe light dark matter in current experiments.

\textbf{\emph{Acknowledgments --}}
We thank the anonymous referee for their insightful comments and helpful suggestions.
We thank Paul Frederik Depta for helpful comments on the numerical approach. TH acknowledges support by the IMPRS-PTFS. MS thanks the Galileo Galilei Institute for Theoretical Physics for the hospitality and the INFN for partial support during the completion of this work.

\bibliography{bibliography}

\end{document}